\documentstyle[aps,epsf]{revtex}

\newcommand{\disregard}[1]{}

\newcommand{\be}{\begin{equation}}
\newcommand{\ee}{\end{equation}}
\newcommand{\ba}{\begin{array}}
\newcommand{\ea}{\end{array}}
\newcommand{\bn}{\begin{eqnarray}}
\newcommand{\en}{\end{eqnarray}}
\newcommand{\bnl}{\begin{mathletters}\begin{eqnarray}}
\newcommand{\enl}{\end{eqnarray}\end{mathletters}}
\newcommand{\bml}{\begin{mathletters}}
\newcommand{\eml}{\end{mathletters}}
\newcommand{\bc}{\begin{center}}
\newcommand{\ec}{\end{center}}
\newcommand{\bi}{\begin{itemize}}
\newcommand{\ei}{\end{itemize}}

\newcommand{\bnll}[1]{\begin{mathletters}\label{#1}\begin{eqnarray}}
\newcommand{\enll}{\end{eqnarray}\end{mathletters}}

\tighten
\begin{document}

\input{epsf}

\draft
\twocolumn[\columnwidth\textwidth\csname@twocolumnfalse\endcsname
\title{Multifragmentation of non-spherical nuclei}

\author{A. Le F\`{e}vre$^{\dagger}$, M. P{\l}oszajczak$^{\dagger}$
and V.D. Toneev$^{\ddagger}$ }

\address{{$\dagger$}
Grand Acc\'{e}l\'{e}rateur National d'Ions Lourds
(GANIL), CEA/DSM -- CNRS/IN2P3, BP 5027, \\
F-14076 Caen Cedex 05, France
\newline           {$\ddagger$}
Bogoliubov Laboratory of Theoretical Physics, Joint
Institute for Nuclear Research, Dubna, 141980 Moscow Region, Russia}

\maketitle

\begin{abstract}
Influence of the  shape of thermalized source on
various characteristics of multifragmentation process
as well as its interplay with
effects of the angular momentum and the collective expansion
are studied for the first time and the most pertinent
variables are proposed. The analysis is based on the
extension of the statistical microcanonical multifragmentation
model.
\end{abstract}


\addvspace{5mm}]

\narrowtext

In studying the multifragmentation process, a large range
of incident energies, changing by about four orders in magnitude, was
covered and various types of projectiles, from proton till  heaviest
available ions, were probed. The reaction mechanism is often
considered in terms of two-step scenario where the first, off-equilibrium and
dynamical step results in the formation of thermalized source
which then, in the second step, decays statistically into light particles
and intermediate-mass fragments (IMF's). Assuming that
the thermal equilibrium is attained, various statistical
multifragmentation models were employed for the second step
(see \cite{gross90,gross97,bond96} and references quoted therein). These models
were so successful in providing an understanding of basic aspects
of the multifragmentation process that the deviations between
their predictions and the experimental data have been often taken as an
indication for dynamical effects in the multifragmentation.
This kind of simplistic 'cause - effect' interpretation may
however be misleading due
to several oversimplifying assumptions in the
statistical calculations, such as , {\it e.g.},
the spherical shape of the thermalized source.
Indeed, one expects that the spherical shape can be
perturbed during the dynamical phase and
 the density evolution has both compressed and rarefied zones which can give
rise to a rather complicated source forms \cite{GIT76,B92}.
Perhaps more important are the
angular momentum induced shape instabilities
\cite{swiatecki,fpj} which may cause large fluctuations
of both the Coulomb barrier
and the surface energy even for moderately high angular
momenta ($L \sim 40 \hbar$). Moreover, at high excitations,
not only the quadrupole stiffness
becomes small but also the fission saddle point
moves towards larger elongations and smaller neck cross-sections \cite{fpj},
giving rise to some 'neck effects' \cite{swiatecki,fpj}.
Hence, before discussing dynamical effects in the multifragmentation decay,
one should study the effects of different shapes in the freeze-out
configuration.
In this paper, the non-spherical fragmenting source is considered within the
statistical model and the observables sensitive to the shape of this source are
searched for.

Our statistical consideration is based on the
MMMC method of the Berlin group \cite{gross90}.
In the MMMC method, one calculates all accessible
states equally populated in the decay of thermalized system into N
fragments.  The microscopic thermodynamics used here describes the
dependence of the volume of 6N-dimensional phase space on globally
conserved quantities (energy, mass, charge, ...) and external constraints
 (like the spatial volume) to be defined by the first stage of the reaction.
Within the microcanonical ensemble method, an explicit treatment
of the fragment
positions in the occupied spatial volume allows for a
direct extension of the MMMC
code \cite{gross90} to the case of non-spherical shapes.
Here, the source deformation is
considered as an additional external constraint.
Main results of our paper  will be given for the source
described as an axial ellipsoid : $(x/R_x)^2+(y/R_y)^2+(z/R_z)^2=1$,
with $R_x=R_y\neq R_z$. We assume that the freeze-out density of
deformed system
is the same as that of a spherical system with the radius
$R_{sys}=(R_xR_yR_z)^{1/3}$, {\it i.e.},
the volume of deformed system is conserved. This
condition changes neither the  pass scheme nor
the weight $w_r$ due to the accessible
volume of the fragments in the Metropolis scheme of calculations
\cite{gross90}.  On the other hand, it means that the ellipsoidal source shape
depends on one additional parameter : the ratio of ellipsoid axes
${\cal R} = R_x/R_z$.  The ratio ${\cal R}<1$ corresponds to the prolate
form, while for the oblate form one has ${\cal R} > 1$.

An essential feature of non-spherical systems is that the deformation
'costs' some extra energy $E_{def}$ which is proportional
to a change of nuclear surface with respect to the spherical shape.
Since we do not consider shape evolution of the system but rather the
influence of source shape on its thermodynamics, this energy $E_{def}$
will be inaccessible for thermal motion and may be disregarded
in the total energy balance.
However this point should be kept in mind if one tries to
refer to the real values of energy pumped into the system.

The source deformation
will noticeably affect the moment of inertia and,
together with the Coulomb energy which is  calculated exactly for   every
multifragment configuration of non-spherical nucleus, becomes very
important for describing rotating systems. As to the general scheme to
account for the total angular momentum and the calculation of the statistical
weight $w_{pl}$ of the configuration in the rotating frame, we
closely follow Ref. \cite{BG95} to be realized in the available code.
\footnote
{In our version of the MMMC code the program error made in \cite{BG95} is
corrected (see also Ref.  \cite{gross97}).}
For each spatial configuration
of fragments, part of the total energy goes into rotation and hence the
temperature of the system will slightly fluctuate.  We take into account
fluctuations of the moment of inertia arising from fluctuations in the
positions of fragments and light particles. In the calculation of statistical
decay of fragmenting system, the angular velocity of the source is added to
the thermal velocity of each fragment.

In calculating all accessible states within the standard MMMC method, the
source should be averaged with respect to
the spatial orientation of its axes, which is assumed to be
homogeneously distributed in the whole $4\pi$ - solid angle
\cite{gross90}. If the angular momentum of fragmenting system, caused
by the dynamical first step of the reaction,
is strictly conserved then not all
these states will be accessible and a formal averaging over the
$4\pi$ - solid angle will result in the violation of angular momentum
conservation \cite{gross97,BG95}. In the considered reactions, the angular
momentum vector is perpendicular to the reaction plane. So, we disentangle in
the MMMC code the beam direction (the $z$ -  axis)
and the rotation axis (the $x$ - axis). The rotation energy is then :
${\bf L}^{2}/2J_x \equiv {L_x}^2/2J_x$, where
$J_x$ is the rigid-body moment of inertia
with respect to the $x$ axis.  Averaging
over the polar angle $\theta$ is not consistent with the angular momentum
conservation. On the contrary, averaging over $2\pi$ in the angle $\phi$
corresponds to averaging over azimuthal angle of the reaction impact
parameter and should be included. Averaging over
rotation angle $\psi$ around ${\bf L}$ depends on the considered
reaction, namely on the relationship between a rotation time~:
$\tau_{rot}=J_x /L_x$, and a characteristic life-time of the source
$\tau_{c}$. For a system with high angular momentum when
$\tau_{rot}\ll\tau_{c} $, the full averaging in $0\le\psi\le 2\pi$ should be
performed.  In the opposite limit when $\tau_{rot}\gg\tau_{c} $, only states
with $\psi \simeq 0$ are accessible.  Below we shall consider both these
limiting cases.

In the HI collisions, a part of the total energy can be stored in the
compression energy of pre-formed source which during the collective
(isentropic) expansion is transformed
into the kinetic energy of fragments. In a strict thermodynamic sense,
such an expanding system in not in equilibrium. However, in the  quasi-static
expansion, {\it i.e.}, when the time scale involved in the expansion is larger
compared to the equilibration time, the system may be  considered to be
 infinitesimally close to the thermal equilibrium and consistently
 treated likewise
an equilibrated  system under the action of a negative external pressure
whose magnitude is equal to the flow pressure. Such an approach has been
applied for describing the flow effect in multifragmentation within a
quantum-statistical model \cite{PSD96}
giving an estimate of about 20
$MeV/A$ for a maximal applicable flow energy.
Consistent treatment of this effect within the microscopic approach
would require an introduction of proper weight factors in the MMMC code.
Such a work is now in progress \cite{future} . However,
to get some insight into the influence  of collective flow on the
multifragmentation process,  we shall mimic this effect  by simply
adding the blast velocity $v_{\displaystyle b}$ to the thermal
velocity of each particle/fragment
for any event simulated by the Metropolis method. This approximate
procedure does not affect the Metropolis pass scheme in the original code.
Assuming ${\bf v} \sim {\bf r}$, a simplified scaling solution of the
non-relativistic hydrodynamic equations describing the radial expansion of a
spherical source provides the following general form for the radial velocity
profile \cite{BGZ78,CLZ94}:
\begin{eqnarray}
\label{bl} v_{\displaystyle b}(r) = v_0
\left({r}/{R_0} \right)^{\displaystyle \alpha} ~~~ \ .
\end{eqnarray}
Here  $v_0$ and $R_0$ are strength and scale parameters of the
flow respectively, and the power-low profile
is characterized by the exponent $\alpha$ , which commonly is taken
in the interval~: $0.5 \le \alpha \le 2$.
For the non-spherical expansion, the velocity
profile may have a more complicated form and depends on the direction. But,
even in the case of axially symmetric expansion, the scaling
relation (\ref{bl}) was successfully applied for describing the profile of
the transverse expansion velocity \cite{CLZ94,PBM98}. Below it is assumed
that the radial expansion at the freeze-out point is described by eq.
(\ref{bl}).
In the hydrodynamic interpretation, the scaling parameter $R_0$
corresponds to the size of the system at the initial time of the scaling
regime and, therefore, it should be less than the effective radius of
the spherical source at the freeze-out point, {\it i.e.}, $R_0 \le R_{sys}$. The
strength parameter $v_0$ is then equal to the blast velocity at the surface
of this effective sphere.  As can be seen from eq. (\ref{bl}), the
kinetic energy of fragments produced
in the 'interior' will be mainly sensitive to the variation of
the exponent $\alpha$ while the total collective flow energy
may be fitted by different combination of all these three
parameters.  Note that the {\it average} collective energy
of expansion, similarly as the deformation energy,
is not included in the value of the total excitation energy.

As an example, let us  consider the multifragmentation of $^{197}Au$ having the
angular momentum $L=40 \hbar$ and the
excitation energy $6 \ A\cdot MeV$. These parameters correspond to the source
formed in central $Xe+Sn$ collisions at $50 \ A\cdot MeV$
 \cite{INDRA}. All calculations were carried out at the
standard break-up density $\rho \approx \rho_0 /6$, what gives
$R_{sys}=12.8 \ fm$ for the radius of spherical $^{197}Au$ nucleus. We consider
two ellipsoidal shapes characterized by the axis ratio ${\cal
R}=0.6$ (prolate shape) and  ${\cal R}=1/0.6=1.667$
(oblate shape).
We have found that none of the observables related to the fragment-size
distribution is sensitive to the deformation of fragmenting source at such
high excitation energies.
The c.m.s. angular distribution of the largest fragment ,
{\it i.e.}, the fragment with the  largest charge $( Z = Z_{max} )$ ,
is shown in Fig.1.  In the absence of collective expansion,
the angular distribution is isotropic for oblate configurations and has small
forward - backward peaks for prolate configurations
if the  averaging around {\bf L} is performed in the whole available interval
$0\le\psi\le 2\pi$.
For the 'frozen' spatial configuration $( \psi = 0 )$,
the 'deformation effect' is clearly
seen: for the prolate form there are strong forward - backward peaks
, while for the
oblate form the heaviest fragment is predominantly emitted  in the sideward
direction (${\theta}_{cm} = \pi /2$),
like in the scenario of hydrodynamic splashing. This deformation effect is
definitely not due to the angular momentum as can be seen by
comparing Figs. 1a, 1b ($L = 0$) with Figs. 1c, 1d ($L = 40 \hbar$).
The collective expansion ($\alpha = 2$)
enhances the deformation effect. One may notice a strong
enhancement of forward and backward peaks in the prolate case and the
appearance of a strong peak at  ${\theta}_{cm} = \pi /2$ in the oblate case.
Similar features can be seen also in the cumulative angular
distributions of all IMF's but the relative amplitude of the deformation effect
in that case is smaller.

\begin{figure}[t]
\begin{center}
\leavevmode
\epsfxsize=7.5cm
\epsfbox{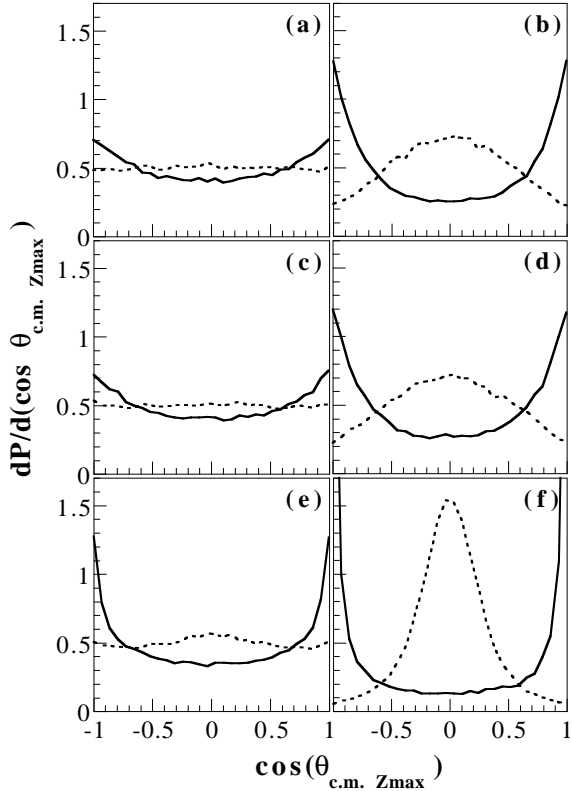}
\end{center}
\caption[C1]{Angular distribution of $Z_{max}$ in the c.m.s. Plots on
the l.h.s. correspond to averaging over the whole available interval
$0\le\psi\le 2\pi$, whereas plots on the r.h.s. correspond to the 'frozen'
configuration  $ \psi = 0 $.
{\bf (a),(b)} : $L=0$, $v_{\displaystyle b}=0$; \
{\bf (c),(d)} : $L=40 \hbar$, $v_{\displaystyle b}=0$; \
{\bf (e),(f)} : $L=40 \hbar$, $v_{\displaystyle b}=0.08 c$, $\alpha$=2 .}
  \label{fig1}
\end{figure}

Large sensitivity to the source shape is expected in the analysis using
global variables on an event-by-event basis \cite{cugnon}.
Here we shall associate the global variables with the momentum tensor :
\begin{eqnarray}
\label{tens}
Q_{\displaystyle ij} = \sum^N_{\nu =1}
{\gamma}^{(\nu )} p^{(\nu )}_i p^{(\nu )}_j ~~~ \ ,
\end{eqnarray}
where $p^{(\nu )}_i$ is the $i$th Cartesian coordinate ($i=1,2,3$) of the
c.m.s. momentum $p^{(\nu )}$ of the fragment $\nu$. The sum

\begin{figure}[thb]
\begin{center}
\leavevmode
\epsfxsize=7.5cm
\epsfbox{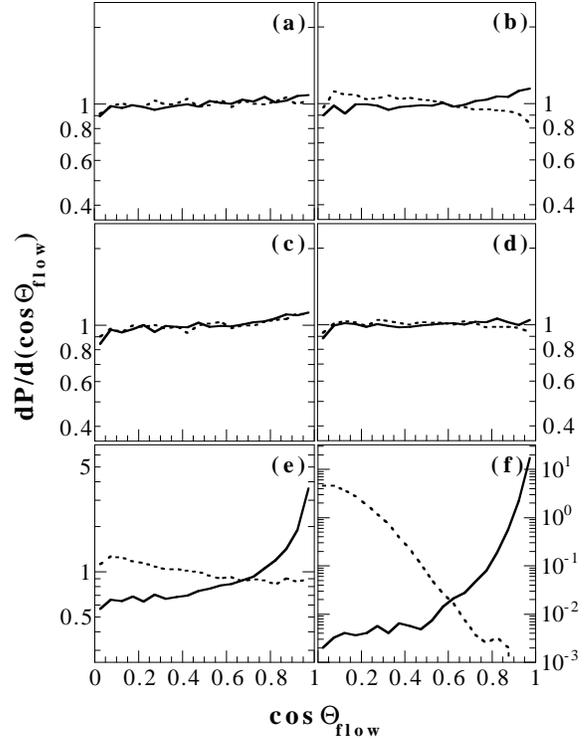}
\end{center}
\caption[C2]{${\Theta}_{flow}$ distribution. The plots on the l.h.s.
correspond to the averaging over the whole available interval
$0\le\psi\le 2\pi$, whereas the plots on the r.h.s. correspond to the 'frozen'
configuration  $ \psi = 0 $.
{\bf (a),(b)} : $L=0$, $v_{\displaystyle b}=0$;
{\bf (c),(d)} : $L=40 \hbar$, $v_{\displaystyle b}=0$;
{\bf (e),(f)} : $L=40 \hbar$, $v_{\displaystyle b}=0.08 c$, $\alpha$=2.}
  \label{fig2}
\end{figure}
\noindent
in (\ref{tens})
is running over all IMF's ($Z \geq 3$). The factor
${\gamma}^{(\nu )}$ depends on the physical interpretation which one wants
to give to the tensor (\ref{tens}). We use $\gamma = 1/2m_{(\nu )}$, where
$m_{(\nu )}$ is the mass of fragment $\nu$. The
tensor $Q_{ij}$ can be represented as an ellipsoid in the momentum
space. The  shape of this ellipsoid can be described by three  axes and
its orientation can be fixed  by three angles  in the
3D-momentum space. This is usually done by referring to the eigenvalues
$0 \le \lambda_1 \le \lambda_2 \le \lambda_3$ ($\lambda_1 + \lambda_2 +
\lambda_3 = 1$)
of the tensor $Q_{ij}$ and to the Euler angles defining the
eigenvectors ${\bf e_1, e_2, e_3}$. From various possible  combination of
these parameters defining global variables \cite{cugnon} ,
we consider here the sphericity :
${\it s} = (3/2) (1-\lambda_3)$, the coplanarity :
${\it c} = ({\sqrt{3}}/2) (\lambda_2-\lambda_1)$, the aplanarity :
${\it a} = (3/2) \lambda_1$,
and the flow angle $\Theta_{flow}$ defined as an angle between $\bf e_1$ and
the $z$-direction (the beam direction) in the c.m.s.

Distribution over the flow angle is presented in Fig. 2.
Different physical situations considered in Fig. 2, are exactly the
same as in Fig. 1. Even though $\Theta_{flow}$ characterizes now all
IMF's rather than the most sensitive largest fragment,
the difference in the source shape manifests
itself already for a vanishing blast velocity $v_{\displaystyle b}=0$.
The whole effect is extremely sensitive to the presence of collective
expansion (see Figs. 2e and 2f) and
is enhanced furthermore for the 'frozen' configuration ($\psi = 0$).
The maximal size of genuine angular momentum effects can be seen by comparing
Figs. 2b ($L = 0$) and 2d ($L = 40 \hbar$). In general one expects that
the $\cos (\Theta_{flow})$ - distribution
for highly central events is uniform \cite{INDRA} .
However this naive expectation may be altered by many effects,
such as the spatial shape of the source, the
non-vanishing collective radial expansion
, the high angular momentum and last but not least the
detection bias.
\begin{figure}[thb]
\begin{center}
\leavevmode
\epsfxsize=5.8cm
\epsfbox{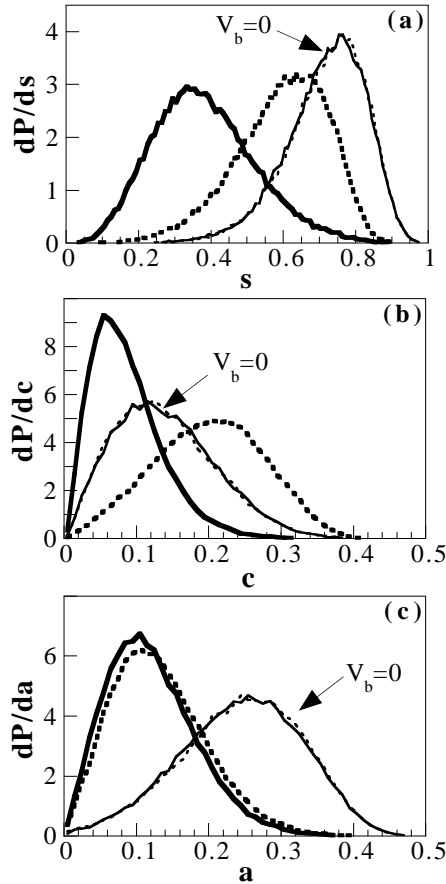}
\end{center}
\caption[C2]{{\bf (a)} Sphericity , {\bf (b)} coplanarity and {\bf (c)}
aplanarity distributions for $L=40 \hbar$. Bold lines correspond to
$v_{\displaystyle b}=0.08 c$, $\alpha$=2. The continuous
lines show prolate ${\cal R}$=0.6 shape whereas dashed lines show oblate
${\cal R}$=1.667 shape.}
  \label{fig3}
\end{figure}

Fig.3 shows the distributions over sphericity $dP/d{\it s}$, coplanarity
$dP/d{\it c}$ and aplanarity $dP/d{\it a}$ for different source deformations.
By construction, these distributions do not depend
on $\psi$-averaging. The curves plotted with bold
lines correspond to non-vanishing collective expansion which reveals
different ellipsoidal source shapes. The reason for this sensitivity
can be seen from (\ref{bl}).
Depending on the source shape, a different number of fragments
can be placed in the two regions : $r<R_0$ and $r>R_0$ (we put everywhere
$R_0=0.7R_{sys}$), in which the expansion acts differently.  For
the same reason, these distributions are insensitive to the
radial expansion for the spherical source.
Particularly interesting are the sphericity and
coplanarity distributions where the evolution of distributions with
$v_{\displaystyle b}$ is clearly different for prolate and oblate
source shapes. Lack of sensitivity of the aplanarity distribution to
the deformation effects and to the expansion happens accidentally for chosen
parameters.
One should also mention that the sensitivity of $dP/d{\it s}$
and $dP/d{\it c}$  distributions to the source deformation  and their
insensitivity to the time-scales involved (the $\psi$ - averaging)
provides interesting and supplementary informations to those
contained in the $Z_{max}$-angular distribution
and in the $\cos (\Theta_{flow})$ - distribution.

 The $Z$-dependence of average kinetic energy $E_k$ of IMF's
for diffferent source shapes and different
$\alpha$-parameters is shown in Fig.4. The calculations have 

\begin{figure}[t]
\begin{center}
\leavevmode
\epsfxsize=7.5cm
\epsfbox{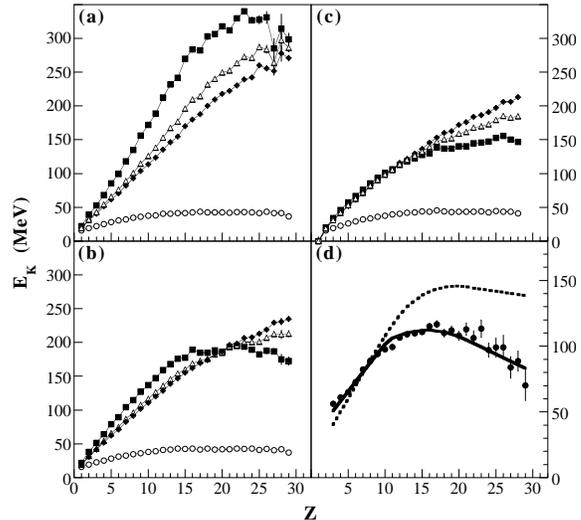}
\end{center}
\caption[C3]{Kinetic energy of fragments is plotted as a function of $Z$.
{\bf (a)} Prolate shape, $L=40 \hbar$. Different curves correspond to
 : $v_{\displaystyle b}=0$ (circles), and : $v_{\displaystyle b}=0.08 c$,
$\alpha =$1/2 (diamonds), 1 (triangles), 2 (squares).
{\bf (b)} The same as in (a) but for the oblate shape.
{\bf (c)} The same as in (a) but for the spherical shape.
{\bf (d)} Comparison between the experimental data and the calculation
($\alpha$=2) for both prolate and oblate freeze-out configurations and the
blast velocities : $v_{\displaystyle b}=0.088 c$ (prolate shape),
$v_{\displaystyle b}=0.063 c$ (oblate shape), chosen as to reproduce the
experimental mean kinetic energy per nucleon of IMF's : $5.2 \pm 0.1 MeV/nucl$. }
  \label{fig4}
\end{figure}
\noindent
been done for the 'frozen' configuration ($\psi$=0).
The collective expansion energy is a dominant contribution to
the average kinetic energy of fragments, and its
value differs noticeably for prolate and oblate source shapes.
It is of interest to note that
%while the average thermal kinetic energy (open
%circles in Fig. 4) increases linearly with $Z$,
the average kinetic energy of
fragments exhibits a flattening or even a maximum at large $Z$ (see , {\it
e.g.}, the curve for $\alpha = 2$).
 The precise position of this maximum depends also on the
deformation (see Figs. 4a and 4b).
An attempt to reduce this large observed kinetic energy to angular
momentum effects results in a value of the average angular momentum
($L\approx 640 \hbar$) which is completely unrealistic for selected
central events \cite{INDRA}. Fig. 4c shows $E_k(Z)$ for the
spherical source shape. Fig. 4d compares results of the present model
with the experimental data
for central $Xe+Sn$ collisions at $50 \ A\cdot MeV$
\cite{INDRA}.  All calculated events have
been filtered with the INDRA software replica, and then
selected with the experimental centrality
condition : complete events ({\it i.e.}, more than 80\%
of the total charge and momentum is
detected) and $\Theta_{flow} \ge\ {\pi}/3$.

In conclusion, external constraints on the shape of
equilibrized fragmenting source have been considered
within the extended
MMMC method. Due to the  change in the Coulomb energy for deformed freeze-out
configuration, the shape effect is clearly seen in the IMF's angular
distributions ($Z_{max}$- angular distribution) as well as
in the $\Theta_{flow}$ - distribution.
%On the contrary, the
%observables related to the fragment-size (or $Z_{max}$) distribution are quite
%insensitive to the source deformation.
A surprising interplay between effects of non-spherical
freeze-out shapes and the memory effects of
nonequilibrium phase of the reaction, such as the rotation and the collective
expansion of the source, has been demonstrated for the first time.
The influence of shape on rotational properties of the
system is not only reduced to the modification of the
momenta of inertia. The limits on the averaging interval over
the angle $\psi$ about the rotation axis, which are defined by the
appropriate time scales, affect strongly the angular
observables and are able  to enhance strongly the
'shape effect'. These constraints may be important for certain
observables used in experimental procedures of selecting
specific  class of events. Other striking finding is that the
collective expansion allows to disclosure the source shape  in  the analysis
 using global variables as well as
in the study of $Z$-dependence of the average kinetic energy. The latter
observable is independent of $\psi$- averaging but, unfortunately, it is
specified by a poorly known profile function (\ref{bl}).
Nevertheless, the careful analysis with
eq. (\ref{bl}) might shed some light on the problem how different fragments
are situated in the freeze-out configuration.

In the experimental analysis, if the average kinetic energy of fragments
is fixed by an appropriate choice of
$v_{\displaystyle b}, R_0$ for each source deformation ,
then the shape of $E_k(Z)$ contains
information about the exponent $\alpha$ in the parametrization (1) and to the
lesser extend about the deformation of the source. The form of the
$dP/ds$ and $dP/dc$ distributions permits then to find the average deformation
of the fragmenting source. Having fixed the parametrization (1)
%of the collective expansion energy
and the deformation of the source, the analysis of the
angular distribution of $Z_{max}$ and/or the ${\Theta}_{flow}$ distribution
gives an access to the information
about the time-constraints (the limits on the $\psi$-averaging)
in the multifragmentation process.
We have discussed here the freeze-out shape effect for only one
value of the excitation energy.
Certainly, the manifestation of this effect in observables
is energy/angular momentum dependent
%One may expect that at low excitation even
%the fragment-size distributions will be disturbed due to a non-spherical
%source deformation. On the other hand, when we proceed from low to higher
%intermediate energies, the shape itself of the fragmenting source
%will change  because the change of the first step mechanism of heavy ion
%collisions.
and the modifications of the MMMC method, presented in this Letter, open
a promising way to follow up the
shape evolution of equilibrized source in
the broad  range of  incident energies. Alongside
with the observables discussed here, it would be interesting
to study the velocity
correlations between fragments which are sensitive to the source shape at
the freeze-out. Such a work is now in progress \cite{future}.

\acknowledgments

We are grateful to D.H.E. Gross for his encouragement and interest
in this project. We are thankful to O. Shapiro for the implementation of the
original version of MMMC code. We thank also G. Auger, A. Chbihi and
J.-P. Wieleczko for their  interest in this work. This work has been supported
by the CNRS-JINR Dubna agreement No 98--28.

\end{document}